# Impliance: A Next Generation Information Management Appliance

Bishwaranjan Bhattacharjee[2]   Vuk Ercegovac[1]   Joseph Glider[1]   Richard Golding[1]
Guy Lohman[1]   Volker Markl[1]   Hamid Pirahesh[1]   Jun Rao[1]
Robert Rees[1]   Frederick Reiss[1]   Eugene Shekita[1]   Garret Swart[1]

{bhatta,vercego,gliderj,rgolding,lohman,marklv,pirahesh,junrao,rees,freiss,shekita, gswart}@us.ibm.com

[1]IBM Almaden Research Center            [2]IBM Watson Research Center

## ABSTRACT

Though database technology has been remarkably successful in building a large market and adapting to the changes of the last three decades, its impact on the broader market of information management is surprisingly limited. If we were to design an information management system from scratch, based upon today's requirements and hardware capabilities, would it look anything like today's database systems? In this paper, we introduce Impliance, a next-generation information management system consisting of hardware and software components integrated to form an easy-to-administer appliance that can store, retrieve, and analyze all types of structured, semi-structured, and unstructured information. We first summarize the trends that will shape information management for the foreseeable future. Those trends imply three major requirements for Impliance: (1) to be able to store, manage, and uniformly query and transform all data, not just structured records; (2) to be able to scale out as the volume of this data grows; and (3) to be simple and robust in operation. We then describe four key ideas that are uniquely combined in Impliance to address these requirements, namely the ideas of: (a) integrating software and off-the-shelf hardware into a generic information appliance; (b) automatically discovering, organizing, and managing all data – unstructured as well as structured – in a uniform way; (c) achieving scale-out by exploiting simple, massive parallel processing, and (d) virtualizing compute and storage resources to unify, simplify, and streamline the management of Impliance. Impliance is an ambitious, long-term effort to define simpler, more robust, and more scalable information systems for tomorrow's enterprises.

## 1. Introduction

While the relational database industry is unquestionably a business success – generating revenues of over $14 billion in 2005 and managing the mission-critical data of virtually every Fortune 1000 enterprise – its impact on information management as a whole remains remarkably disappointing. Despite concerted efforts in the late 1980s to build "extensible databases", most of the world's data – 80% and growing – is not stored in databases [39]. Though the standardization of SQL spurred a whole new industry of applications based upon its interface, keyword search is today a far more popular retrieval paradigm than SQL. While the last decade has seen tremendous advances in making databases more self-managing and easier to use, users still complain that databases take too long and too much expertise to set up, configure, tune, test, deploy, maintain, and enhance. While databases are arguably the most successful exploiter of parallel architectures, today even the largest deployments rarely exceed a few hundred nodes. And though database architectures have endured several orders of magnitude changes in the relative speeds of hardware over the last 30 years, the many tiers and layers of components in today's information management systems are exceedingly complicated, containing redundant and competing components. It is time to re-examine not only the architecture of database systems as we know them today, but also their place in the overall stack of software employed in modern data-intensive applications. If we were to design an information management system from scratch based upon the anticipated requirements and hardware trends we know now, how would it differ from the products and research prototypes available today?

This paper describes an effort to do just that – to design and build a comprehensive "next-generation" information management system "outside-in", starting with the top-level requirements of enterprise information management, and working backward to define the software and hardware needed to meet those requirements. This process has resulted in a radical new architecture that integrates software and hardware into a high-function and easy-to-manage information management appliance that we call Impliance, currently being designed and prototyped at the IBM Almaden Research Center. Impliance will be capable of storing, retrieving, and analyzing all types of structured, semi-structured, and unstructured information, with low total cost of ownership (TCO), and supporting the needs of small businesses to those of the largest global enterprises.

The requirements of Impliance are motivated by the confluence of several major trends in information management:

- All data types – not just structured homogeneous tables – need to be managed uniformly by tomorrow's information management systems. The requirements for structured, semi-structured, and unstructured information are converging, requiring databases to manage huge amounts of information of various data types (PDF, XML, Text, Audio, and Video) that do not adhere to predefined schemas. Customers increasingly want to search, classify, aggregate, and analyze the content of semi-structured and unstructured data in new ways, and find today's distinct systems for database, content, and files too complicated and non-uniform in such functionality.
- Data volume growth is accelerating. Content data such as documents and digital media, as well as more structured information generated by sensors in large volumes (e.g., RFID streams), are contributing to this trend. Additionally, enterprises are retaining their data on-line longer, to comply with records retention requirements and perform longer-term analysis of business trends. Customers want systems that can seamlessly and scalably expand as an enterprise and its systems' needs grow.
- Total cost of ownership (TCO) is increasingly dominated by labor costs. Hardware and software costs already represent less than half of the cost of most IT systems, and are decreasing rapidly, while labor-related costs such as deployment, maintenance, and problem determination are increasing. Customers want systems that are easy to install, deploy, use, and manage, and that minimize the need for human intervention. Retrofitting layers of autonomic management can lead to increased complexity and difficulty of problem determination. True simplicity, modularity, and low TCO can best be achieved by designing it into the system from inception.
- Enterprises increasingly need to integrate and analyze information from independent "silos" serving different applications, organizations, or geographies in order to support regulatory compliance, mergers and acquisitions, and other business imperatives. The current generation of information integration products gives access to the data but rarely provide a means of querying across silos based on the semantics of the data; instead cross silo analysis requires that the data be transformed into a single preferred schema/format. These deficiencies increase the cost of developing new applications and of maintaining the data; integrating systems *ex post facto* is human-intensive and thus prohibitively expensive.
- Hardware architectures have changed and scaled far more dramatically than software architectures. Current information management software products are largely based on the hardware architectures of several decades ago. While those products have shown remarkable resiliency to order-of-magnitude changes in hardware speeds and capacities, re-examining the software architecture in light of tomorrow's hardware seems overdue. Drastic improvements in hardware – including low-power multi-core blade servers, large memories (but with relatively higher latency to get to), layers of large on-chip caches (with low latency), ultra-dense storage systems, and commodity low-latency networks – provide novel system building blocks that can be exploited better with new software architectures.

Based on the above trends, we identify three important requirements for the next generation information management system: (1) the ability to store and manage all types of data and to perform advanced business analytics and information retrieval on that information; (2) scalability at least an order of magnitude higher than today's database management systems; and (3) extremely low TCO by shifting human brain cycles to machine cycles. Such a system exists neither in the market nor the research lab today, which motivates our vision for Impliance.

Impliance is designed either to be installed at a customer site, or to be run by a third party service provider as a component of an online service. It will be the repository of operational, warehouse, and archival data. While Impliance will support transactions and will be a component in a distributed system, the primary focus of our research is providing analysis and integration of information stored within a single scalable appliance, rather than high volume OLTP or global scaling across a WAN.

We first paint an overall picture of Impliance in Section 2. Then, in Section 3, we describe a few of the key ideas we are applying to address the requirements identified above. Other important issues to be solved are identified in Section 4. We summarize some of the related work in Section 5, and conclude in Section 6.

## 2. The Big Picture

In this section, we give some example use cases indicative of the type of problem Impliance is trying to solve, and a broad overview of the functionality that Impliance provides to satisfy those use cases.

### 2.1 Use Cases

We first illustrate the necessity for Impliance with real business use cases whose requirements are difficult or impossible for today's software products to satisfy. This will serve to motivate the remainder of the paper.

*2.1.1 Exploiting Customer Relationship Management*

Virtually every modern company has at least one center for handling customer phone calls, e-mails, and/or web-page comments, questions, and complaints. Besides maintaining customer satisfaction, these touch points with customers provide an excellent opportunity for selling more products to existing and prospective customers. An alert call center operator may recognize that the solution to a customer problem might be another product the company offers, or might even sense that the caller is open to buying additional products. Today this requires either highly trained and motivated (and hence expensive) operators, or canned sales pitches that are scripted in advance and often not appropriate for that customer. Ideally, the company would capture the customers' words and extract from them what products they know about, might be interested in, and even their opinion of the company's products. By correlating the information extracted from the text of the conversation transcripts with the profile of similar customers who are happy with what they bought, one can provide a customized offer to a customer through a combination of services and products.

*2.1.2 Integrating Content and Data*

Today, enterprises typically manage semi-structured or unstructured information such as text, images, forms, and video using content management products. These products have very limited awareness of the semantics of the content or capabilities to search it, usually restricting search to the content's metadata. Enterprises need the ability to search the actual content and relate it to structured information from other sources. For example, insurance companies looking for fraudulent claims need to find the names of procedures or pharmaceuticals within the text of claim forms or doctors' statements, and relate that to known, structured information about the patient, the provider, the procedure, and/or the pharmaceutical. Forms filled in by claims adjustors with descriptions of vehicle damage and repairs need to be related to images of the damage and the police report of the accident, and compared with reference data from similar accidents to determine if the repair estimate is excessive. Today, the rules that relate the structured and unstructured data are diffused into the logic of dozens of applications and the results surfaced in hundreds of reports. By bringing the data together, the analysis can be systematized, executed more efficiently, and the results made available for use as the basis for further analysis and online retrieval.

*2.1.3 Legal Compliance*

Whenever an enterprise is involved in legal actions with another enterprise, the court-ordered discovery process often requires each litigant to locate and preserve broad classes of information that may be pertinent to the litigation. Much of the relevant information is semi-structured or unstructured – such as e-mail, contracts, meeting transcripts, spreadsheets, patents, SEC filings – and some of the information is highly structured. Just finding this information, much less ensuring its preservation, can be exceedingly expensive and error-prone when the data is in diverse types of repositories and in different formats. Furthermore, the relevance of data may be due to indirect contractual relationships such as partnerships with other enterprises and may require determining the transitive closure of relationships extracted from the content. Proactive auditing is essential to detect legally dubious behavior before it reaches the front page of the newspaper.

## 2.2 Functionality Overview

The use cases above provide a small taste of a broad class of informational needs not adequately met by today's diverse set of information management products, each with its own capabilities, data model, externalized APIs, management requirements, idiosyncrasies, and compatibilities. Maintaining organizations with the requisite skills for each product is an expensive headache. What customers want is greater integration of these diverse data sources, and that is the goal of Impliance.

Specifically, we have identified four major areas of functionality that span information management and that Impliance needs to unify:

1. Semantics. What the data actually means is provided in databases by humans via its logical schema, but in other data repositories the semantics are unknown or implicitly contained in applications, and can only be discovered or inferred from the data automatically through text analytics and annotation, image recognition algorithms, etc. Even tagged data such as XML may not have sufficient semantic information behind the tags to properly relate it, e.g. the units of measurement, how that data was collected, or what it includes and excludes. Having semantic information is a prerequisite to effectively performing the remaining three functions.

2. Search/query. In many cases the data is too voluminous or disparate to process in its entirety, so each use begins by obtaining a subset of the data that meets certain conditions on its content, context or metadata. For structured data, this is done with SQL; for XML it is done with XQuery; and for other data types it is typically done with keyword search, naming conventions, or by providing a unique identifier that is magically known by the requestor. The more semantic information we can extract from the data, the more we can improve the utility of this search.

3. Composition. Relating objects to each other and composing them into new objects creates new information that is at the heart of the value proposition of information management. In databases, this starts with joins relating objects, but it can involve much more. Composition includes integrating items from heterogeneous data "silos" to form higher-level objects, or "quick and dirty" mash-ups that beneficially merge public web data sources with each other or with enterprise data sources. Again, this integration can be done far more effectively (and correctly!) if the semantics of the inputs are more precisely defined.

4. Aggregation. To be consumable by humans, large bodies of data must be reduced through aggregation along various dimensions, to discover higher-level models, trends, and exceptions that facilitate business decisions. Aggregation is fundamental to today's Business Intelligence and On-Line Analytic Processing (OLAP), data mining, and visualization. But aggregating entities embedded in unstructured information, such as text, is extremely hard to do today. As with composition and search, aggregation presupposes sufficient semantic information that we don't aggregate oranges and orangutans. It makes no sense to average phone numbers or salaries paid in different currencies, to double-count revenues contained in diverse sources (e.g., e-mail and a spreadsheet), or to count the total number of completely unrelated entities.

Impliance provides these four basic functionalities for all types of data. Impliance functions as a stewing pot, a repository into which all of your favorite data can be thrown, with no preparation and in any type, schema, or format. Although you can selectively ladle out the unchanged initial ingredients immediately, after simmering on low for a while, you can also fill your bowl with an information Jambalaya: data that has been extracted through an enhanced retrieval interface, spiced with additional semantics through the discovery processes, melded together into interesting new flavors through composition, and boiled down with aggregation to produce an enriched stew of information.

As illustrated in Figure 1, the data infused into Impliance is mapped from its initial format to a uniform data model as described in Section 3.2. Once in this uniform format, the query processing engine can store it and execute queries over it as described in Section 3.3 and 3.4. The discovery process executes queries over the data and uses the results to derive annotations that are added to the data. The end user uses an interactive retrieval interface to find the desired information, optionally making use of the annotations added by the discovery process and domain knowledge built into the repository or the retrieval interface. Note that the query processing engine does not "understand" the annotations; instead, it supports a mixed data/meta-data model that relies on smart query construction by the retrieval interface and discovery engine to decide how to make use of the annotations.

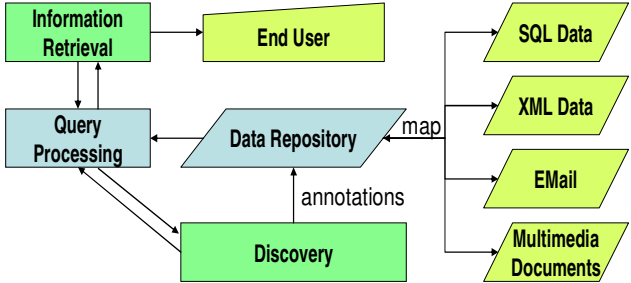

**Figure 1: Impliance Overview**

## 3. Main Ideas

This section summarizes the key new ideas in Impliance. While each alone may not be totally new, together they constitute a revolutionary approach to information management. Each of the ideas addresses issues related to one or more of the requirements given in Section 1.

### 3.1 The Appliance Model

First and foremost, Impliance is an <u>appliance</u> that pre-packages storage, servers, and software into a turn-key information management system that is operational "out of the box". Unlike traditional database machines [10], which advocated special-purpose hardware that did not enjoy the economies of scale of general-purpose chips, Impliance exploits more cost-effective off-the-shelf, consumer-based storage, CPU, and inter-connect technologies configured specifically for information management tasks. This does not mean that one size fits all. To adapt to differing workload requirements and to "ride the technology wave" as component technologies improve at different rates, the number and type of nodes for storage, computation, and communication may differ and may evolve over time. The necessary software is pre-installed, automatically detecting which hardware components are available and re-configuring itself if there are changes. The pre-installation and pre-configuration of the system significantly reduces the "time to value" (TTV), that is, the time between the decision to purchase a system and when its deployment actually realizes benefit for the enterprise, the top priority of data warehouse managers, according to a recent Winter Corp. survey [39]. Using technology widely employed today for user applications, Impliance software upgrades are automatically pushed to the nodes and installed automatically according to user-modifiable policies that balance the performance and availability impact of doing the upgrade with the hope for security and reliability gains.

Another benefit of the appliance model is better integration of different software components. Such integration first enables better collaboration among those components. For example, today's database systems still have difficulty communicating to the underlying operating system and storage units the context of their reference patterns, due to the absence of proper interfaces in general-purpose systems that support anything beyond generic requests to read and write pages. As a result, operating systems and storage units end up trying to deduce the context of page requests by mining reference patterns, often pre-fetching pages that go unreferenced and thrashing their hypothesized pattern when the database queries change subtly, even though the database knows full well from its access plan whether those references are due to a sequential table scan or an un-clustered index scan, and when it can pre-fetch to advantage. Because Impliance controls its entire software environment and exposes to the user only the top-most information management services, extensions to interfaces of operating and storage systems can be made without affecting user applications [36]. Tight integration among layers of software can also improve efficiency by moving functionality to the best component. For instance, higher-level functionality such as aggregation and predicate application can be more easily "pushed down" closer to the storage for early data reduction. Another good example for pushing down logic is compression and encryption. The former is crucial for dealing with large amounts of data, and the latter might be required for security reasons. In contrast to the approach used in [50], the push-down logic is implemented in the software component of a storage unit, and thus can be deployed on any type of commodity hardware. To significantly accelerate the development of Impliance, its implementation will draw code and algorithms from the abundance of mature open-source and commercial software in operating systems, databases, full-text indexes, and content management systems, when appropriate, but modified to exploit the data-friendly confines of the appliance environment to realize smoother integration among software components and more uniform systems management.

### 3.2 Uniform Management of All Data

Today's systems manage different types of data in very different ways, with very different user interfaces, despite the common requirement to reliably store, accurately search the content of, and rapidly retrieve both data and metadata about that content. Databases have typically been limited to managing highly-structured data with a common format and relatively small attributes, which conforms nicely to the tables of relational database systems. Despite the best intentions of extensible object-relational [22][37] and object-oriented [38] databases, other types of data have been largely relegated to unsearchable binary large objects (BLOBs). Only in the last few years have database systems begun to treat XML as a native type, permitting declarative querying of the contents and structure of XML documents whose schemas may vary from one row to the next [8][30][34]. The repository of choice for most semi-structured "content" (documents, forms, images, video,…) in most enterprises is still content managers [44][42], which typically use BLOBs or a file system to store the content, and database systems to manage the metadata (catalog) of that content. Hence searching and querying are limited to the metadata about that content, such as its source, date of entry, and a few other critical descriptors. A recent Java content repository standard (JSR 170) [48] allows querying of metadata. However, all metadata must match a predefined JSR schema; hence schema chaos (diversity) is not supported, as in a conventional content management system. Lastly, the ultra-simple "bag of bytes" model of file systems provides a "repository of last resort" that can manage unstructured as well as structured data, but without the powerful querying capability (e.g., joins and aggregations) we take for granted in databases.

Impliance unifies the management of all data under one umbrella, providing interfaces to search structured and unstructured content and metadata alike. Similar to traditional Information Retrieval (IR) systems, we view the input to Impliance as a collection of documents (or objects), each of which may have its own schema. This way, all types of data can be incorporated into Impliance. When data is persisted, it is first persisted in Impliance's native format. It may subsequently be transformed into different formats or combined with other documents, based on common contents or access patterns, and stored in one or more transformed states that are easier to process. Impliance treats each such new version of a data item as immutable. This versioning obviates the need to update all replicas of a document consistently and synchronously, reducing the problem to determining whether any replica has the most recent version, if that matters to the application.

Impliance automatically indexes each document by its values as well as its structures (e.g., every path in the document) for efficient keyword and structural search. Unlike traditional database systems, this indexing need not take place as part of the same transaction that infused that document initially. Impliance even goes beyond a conventional IR system, however, by permitting automated information discovery at any time, not just at data loading time. All data entering into Impliance will also go through a number of asynchronous analysis phases. Impliance will optionally piggyback data mining algorithms on discovery passes, or perform both opportunistically on any page retrieved into the buffer for other reasons, to more proactively discover trends and exceptions in the data.

Much of the previous research in information discovery can be applied here. First, additional metadata will be extracted for each document by running different kinds of annotators [40][53]. This will identify not only entities such as person names and locations, but also relationships among them. Second, using schema mapping technologies [9][31], structures from different sources can be consolidated. Thus, customer purchase orders can all be searched together, whether they are ingested into Impliance via e-mail, a spreadsheet, a Microsoft Word document, a relational row, or other formats. Finally, additional relationships across documents can be identified by running various analyses on all pairs of documents (conceptually). One such example is entity relationship resolution [28]. As another example, a purchase order can be identified to reference several master data records, including detailed information about a certain customer and product. Discovered relationships can be stored as join indexes and utilized at query time. Such a process of self-organizing heterogeneous data not only facilitates more accurate searching, but also meaningful aggregation, which is at the heart of business analytics. Yet in Impliance the user need not specify this structure *a priori* through a lengthy – and limiting! – process to identify the best logical and physical schemas, thereby further reducing the time to value.

Consider, for example, the insertion of a relational row into Impliance. The row is first mapped into the Impliance document model and persisted. The row can immediately be queried by SQL and retrieved without change. In the background, however, the row is annotated by annotators that have expressed an interest in this type of data and whose results are in demand by annotation consumers. The annotators create new annotation documents that refer to the initial row document, and contain information extracted from the row or additional references forming an association between this document and others already stored in Impliance. At that point the new document and its annotations can be exploited by native Impliance IR-like search components. These derived annotations and associations may themselves be exposed to SQL applications through system-supplied views that map the native data types back into relational rows. Exploiting views in this way (Figure 2) facilitates adding new functionality to existing applications without having to rewrite the entire application to use new APIs.

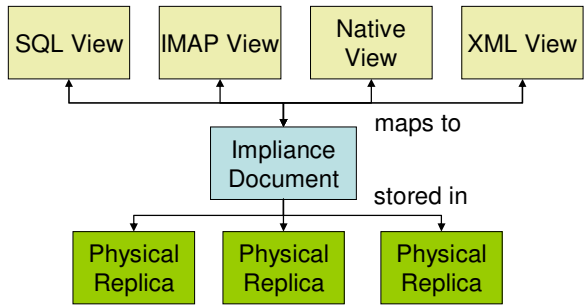

**Figure 2: Impliance Data Modeling**

*3.2.1 Query Interfaces*

Impliance will support two query interfaces. The first one is keyword-driven search, and can immediately be used out of the box. The conventional keyword search interface is simple, but weak for business analytics. Multi-faceted search [27], or guided search, which has been leveraged by search engines such as Endeca [43], iPhrase [47] and Solr [35], provides more analytical functions such as drill-down and drill-across of the search results, while at the same time masking schema complexity from the user through interactive navigational links. We envision an interface for Impliance that extends the concept of faceted search by incorporating more sophisticated analytical capabilities than just counting entities in one dimension, via a sequence of processes that guide the user. For instance, such capabilities will include some flavor of joins and aggregates in traditional relational terms, as well as certain mining operations. Semantic information discovered since data loading time will be exploited for query answering as well. In essence, we are proposing a query interface that brings together keyword search, faceted search, and aspects from traditional OLAP in order to accommodate imprecise search and exploit rich structure that is discovered over time.

The second, more powerful query interface supported by Impliance is intended for building applications that access information through more structured search. We are still in an early phase of understanding the requirements of such a query language. Our preliminary study suggests that it will be a graph-based, web semantics-oriented query interface [6], and designed to facilitate application-level user interaction for query refinement. For example, given two pieces of data, we should be able to ask how they are connected. Traditional structured query languages such as SQL and XQuery can be mapped to this new query interface.

### 3.3 Simple, Massive Parallelism for Query Processing

A typical Impliance installation will consist of several instances of Impliance deployed in geographically separated locations for disaster recovery as well as load balancing. Physically, Impliance consists of a set of devices deployed in one or more racks, connected to each other with a high capacity network and connected to external networks and power in many places. However, its clients and administrator see a single system image. We now focus on the internal design of a single Impliance instance.

In order for a single instance of Impliance to be able to scale from a one-terabyte small business to a multi-petabyte enterprise, the storage and data processing capabilities must be scalable over three orders of magnitude and a wide variety of workloads. To do this while minimizing the resource and, more importantly, the administrative requirements, Impliance needs an efficient way of organizing the storage, computations, and the topology of the underlying hardware. We introduce a novel systems topology and processing architecture that scales in a way that is appropriate to the workload. Each Impliance instance consists of a number of nodes, topologically differentiated into three flavors, each optimized for a particular style of computation based on their connectivity, but each supporting the same execution environment (depicted in Figure 3). These node types correspond to the most popular distributed computing paradigms in use today, but are novel in their use in tightly-coupled combination.

- Data nodes have direct ownership of a subset of the persistent storage, and are the most efficient when performing operations on that storage. Data nodes are sized to balance their computing capability and their I/O bandwidth, but they can be a bottleneck if the data stored on a data node is heavily used.
- Grid nodes perform analytic computations. They may be pulled into a "work crew" to perform long- or short-term operations, and have no long-term state. Grid nodes may offer specialized computing capabilities, such as a hardware accelerator, and have the lowest cost per cycle.
- Cluster nodes are responsible for making consistent locking and caching decisions on data within data consistency groups. Such nodes are good at scalably performing many small consistent updates over a large set of data, but being a part of a consistency group requires overhead for heart-beats and for reacting to nodes joining or leaving the group.

For example, a query can be parallelized by performing full-text index search on a set of data nodes, which then send the reduced data to a set of grid nodes for joining, sorting, and group-wise aggregation, the results of which are sent to a set of cluster nodes to drive a set of updates. For better resource utilization, each operation could be executed on any of the node types. However, the scheduler assigns operators to compute nodes based on which operators execute more efficiently – or with greater

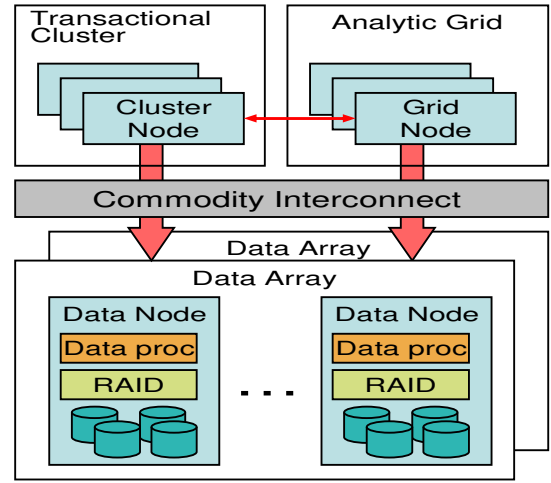

**Figure 3: An Impliance Cluster**

scalability – on a particular node type, the communication pattern of the operator and the availability of resources within the system. Because Impliance is an appliance, it knows about and can model all of its constituent operators and compute nodes, so it can make informed scheduling decisions. Compared with conventional massive parallel systems [16], this architecture makes it easy for Impliance to scale data and processing independently: Add more data nodes to provide additional data capacity or throughput; add more computing nodes to support additional users or applications [26]. As new special-purpose blade hardware becomes available, it can be assigned to run software components that can exploit its capabilities. For example, Cell Blades [52] have excellent vector processing and capabilities and are promising for tasks like compression, encryption and image analysis. Note that both querying and proactive discovery share the same system infrastructure, but exploit it in different ways.

Given a query, Impliance determines an execution plan to deploy in the above infrastructure. Instead of implementing a full-fledged cost-based optimizer as a conventional database system does, we propose to build a simple planner that allows only a few limited choices of the underlying physical operators. Such a planner is desirable because it offers predictable performance (as opposed to optimal performance) and obviates the need for maintaining complex statistics, both of which help reduce a significant amount of TCO present in managing a traditional database system. We believe that such a simple planner is feasible for several reasons. First, we are willing to use more hardware resources to reduce the number of different physical operators that we have to implement and choose from. Second, we can take advantage of the characteristics of the retrieval interface and the underlying data. For instance, given a keyword-search interface that requires only the top-k results, indexed nested-loop joins may always be the preferred join method. As another example, if data is imprecise by nature (e.g., because of fuzzy annotation), the process of summarizing query results can be simplified, since exact answers are not meaningful anyway. Third, the field of adaptive query processing [5][29][32] has advanced significantly over the past six years, and we can borrow and extend some of the techniques to make query operators self-adaptable at runtime.

In addition to interactive queries, an Impliance cluster will run a series of continuous background tasks, such as extracting structured annotations from existing documents. Annotation extraction requires the capabilities of all three node types. Data nodes perform *intra-document analyses:* tasks like entity extraction and sentiment detection within a single document. The output of intra-document analyses may be fed to grid nodes for *inter-document analyses* that identify relationships spanning multiple documents. Finally, cluster nodes are responsible for persisting newly extracted structures and relationships reliably and consistently.

To speedup search queries, Impliance may embed a full-text indexer such as Lucene [24] and Indri [46]. However, we will need to extend those off-the-shelf libraries to meet new requirements in Impliance. For example, for certain kinds of documents, the text indexer has to support hierarchies natively. Techniques of indexing semi-structured data such as XML [18] can be adopted here. Also, for efficient multi-faceted search, it may be useful to extend the payload in the index with additional structured data. Finally, it is important to be able to incrementally maintain the index, especially when structured annotations are added continuously.

### 3.4 Compute and Storage Resource Virtualization

In order to achieve its scalability goals, enterprise deployments of Impliance will be organized as potentially thousands of interconnected nodes constructed from commodity hardware components. Realistically, we expect these nodes to be heterogeneous in composition over time, especially as hardware is upgraded over time. Each node may consist of varying amounts of processing, memory, and disk storage capability. Nor do we assume uniform communication latency between any two nodes; rather, we assume that compute and storage resources are available as collections of smaller groups of tightly-coupled clusters.

In order to unify and simplify its management, Impliance will virtualize this diverse set of compute and storage resources by introducing the notion of a resource group: a group of tightly-coupled nodes (together with their attached storage) that can be assigned the role of cluster, grid, or data storage service, as described in Section 3.3. In order to scale to very large numbers of nodes, we organize and manage these resource groups in a hierarchical fashion [1][20].

The cost-effective autonomic management of these resource groups is a key factor in meeting our goal of reducing both the TCO per byte of data stored, as well as the time-to-value. There are two aspects to this management: execution management and storage management.

Execution management is the task of assigning parts of any task to resource groups, depending on the availability of those groups' resources. For example, it may make sense to execute part of a query such as predicate application on the storage nodes, in order to obtain highest performance and avoid affecting grid nodes. At other times, however, the storage nodes may be too busy serving data for many different analyses, and so moving more work to grid nodes will be preferred. Execution management also includes scheduling prioritized tasks, i.e., managing queues of long-running analysis tasks and properly interleaving these analysis tasks with the execution of queries with more stringent response-time requirements.

Storage management is the task of determining how and where to store the system's data, including how much to replicate the data for reliability. Some data, especially data users have added, will require high reliability, and some will require the kind of regulatory protection mandated by Sarbanes-Oxley. Other data can be re-created with varying amounts of effort, such as data derived by analytics or redundant versions of base data to enhance access or reliability

(e.g., materialized views, indexes, and replicas). Some data requires extra protection, not because it is itself important, but because access to other important data relies on it.

In today's information systems, the administrator has plenty of knobs to turn, at the storage level (choosing the RAID level), in the file system (replication and data placement), information manager (caching and block size), and cluster manager (priority and resource assignment) to implement a desired service level. Our goal is for Impliance to tune all these resources autonomically, perhaps not as well as the best administrator can tune the most configurable products, but instead to utilize resources well enough to deliver cost-effective performance.

For scalability, we envision a hierarchical treatment to both the execution and storage management tasks. At the bottom of the hierarchy are resource groups that provide a pool of compute and storage resources that meet a high-level specification of performance, capacity, and reliability, and act together in the role of cluster service, grid service, or data storage service. Each group manages itself autonomously by scheduling work locally with respect to these service levels and ensuring that the proper amount of resource is available.

Higher in the hierarchy are components that perform macro-level scheduling of jobs to resources groups, as well as components that act as brokers for facilitating the transfer of resources between groups. For example, when a group reports the failure or loss of a resource, it can contact a broker to help it acquire resources from some other group that is willing to relinquish them. Similarly, when new compute or storage resources are added, brokers offer these resources to the groups that will make best use of them.

## 4. Other Issues

Security and versioning are important to Impliance, but are not the initial focus of our research. Since Impliance is designed for enterprise information management, it needs to support policy-driven access controls in such a way that information is provided to the right people, and only to the right people. Another aspect of security is monitoring and auditing. Impliance should be able to trace the lineage [13] of a piece of data as well as queries that have accessed it [2]. Security is the focus of recent enterprise search offerings from both Oracle [51] and IBM [45].

Another important issue is versioning. Because of auditing requirements and the abundance of low-cost storage capacity [21], Impliance does not update data in-place. Instead, changes are implemented as the addition of a new version. We are still investigating whether we should only support a simple sequential versioning primitive and let various other versioning schemes be built on top of it, or directly support more complex ones, allowing branching and merging of versions, as in typical source-code management systems.

## 5. Related Work

Because of its ambitious scope, Impliance builds on a wide base of published research literature. Integrating Information Retrieval (IR) and database systems is currently a hot topic. For example, [3] tries to incorporate keyword search into conventional relational database systems for simplified data exploration, while [12] outlines the benefits and challenges of integrating database and IR technologies. The main research topics identified include user-defined ranking functions and a rank-aware optimizer and algebra. More recently, [25] proposes using "dataspace" to handle all types of data. The basic ideas include schema integration on demand, supporting multiple data models, and exploiting human attention. [4] advocates a tight integration of the functionality of databases and content management systems. The Infosphere project [9] aims to develop a next-generation integrated database management system that incorporates advanced techniques in information integration, content management, and data warehouse systems. While a lot of the ideas are complementary to Impliance, none of them uses the appliance model or shares the focus on the scalability and TCO issues that Impliance does.

Internet information providers such as Google, Yahoo and MSN need to store, manage and retrieve all Web data, which has a size in the order of tens of terabytes in 2005. They have all chosen to build certain kinds of information management infrastructure in-house. For example, the Google File System [19] is a scalable distributed file system and provides fault tolerance while running on inexpensive commodity hardware. Sitting on top of Google File are a highly scalable programming model, called MapReduce [15] and a programming language Sawzall [33]. Using MapReduce and Sawzall together, developers can write and schedule jobs that perform sophisticated analysis on a large data collection. The developers only need to design the logic of a job, and the infrastructure handles parallelism, fault tolerance, and workload balancing. More recently, Google implemented Big Table [11], a distributed storage system for managing structured data that scales to terabytes. Similar to the relational model, Big Table provides an abstraction that simplifies the logic design of a job. Such infrastructures exist in Yahoo and MSN as well. For example, Yahoo has been heavily involved in an open source project, Hadoop [23], which implements a distributed file system and MapReduce in Java. MSN uses Dryad [17], a programming model for a distributed environment (similar to MapReduce). We note that IBM also offers an information integration product DataStage [14], which supports a distributed programming model for ETL (Extract, Transform and Load).

Industry has not been idle in the data appliance market. Netezza [50] and Datallegro [41] both offer appliances for business intelligence applications on relational data. Similar to Impliance, they integrate the hardware and software to reduce the time to value, and rely on simple, massive parallelism to reduce TCO. While Netezza adopts the idea of database machines [10] by using proprietary disk controllers for faster data access, Datallegro uses general purpose hardware, as does Impliance. However, Impliance is intended for managing all types of data, not just relational data, and is designed to scale larger. Google Base [7] provides a service that allows various types of data to be published in a simple way, and to be searchable using a keyword-driven, facet-like interface. In comparison, Impliance focuses more on proactive information discovery, richer business analytics, and data management.

Traditional middleware providers also recognize the need to support heterogeneous types of data within an enterprise. For example, both Oracle Secure Enterprise Search (OSES) [51] and IBM Websphere Information Integrator Omnifind Edition [45] enable many different types of data to be crawled and searched in a secure way. Again, the interfaces that they support are not as advanced as Impliance. In Figure 4 we summarize qualitatively the differences between Impliance and other systems, along the dimensions of scalability, TCO, and modeling and querying power. On the low side of the data richness axis, NetApp [49] provides file system appliances that scale up to 500 TB.

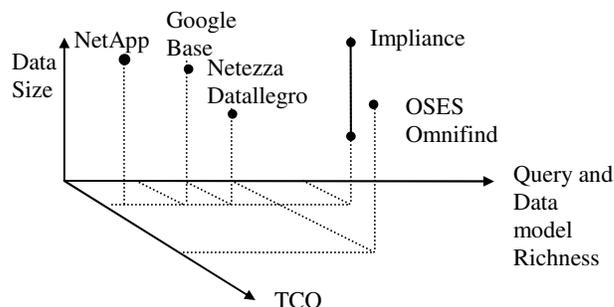

**Figure 4: Comparison between Impliance and Others**

## 6. Conclusions and Future Directions

This paper described Impliance, a next-generation information management appliance that is currently under design at IBM Almaden Research Center. The goal for Impliance is to become a high-function, easy-to-administer system that is capable of storing, retrieving, and analyzing all types of structured, semi-structured, and unstructured information, has low total cost of ownership (TCO), and scales out from small businesses to the largest global enterprises. Impliance can be either delivered directly within an enterprise, or be used to build an online service serving multiple enterprises. To achieve our ambitious goal, we employ four key ideas in the design of Impliance: using the appliance model; managing all data using a single data model; simple, massive parallelism; and virtualized management of compute and storage resources. We plan to collaborate extensively with people both inside and outside of IBM to invent and incorporate the best technologies needed for Impliance's data representation, discovery, retrieval, management, etc.